ADB Economics Working Paper Series

# The Role of Family Support in the Well-Being of Older People: Evidence from Malaysia and Viet Nam

Yana van der Meulen Rodgers,
Joseph E. Zveglich, Jr., Khadija Ali,
and Hanna Xue

No. 730 | June 2024

Yana van der Meulen Rodgers (yana.rodgers@rutgers.edu)
is a professor of Labor Studies and Employment
Relations and faculty director of the Center for Women
and Work and Hanna Xue (hmx1@smlr.rutgers.edu) is a
PhD candidate at Rutgers University. Joseph E. Zveglich,
Jr. (jezveglich@adb.org) is the deputy chief economist
at the Economic Research and Development Impact
Department and Khadija Ali (kali@adb.org) is an
operations analyst at the Central and West Asia
Department, Asian Development Bank.

ASIAN DEVELOPMENT BANK

# ABSTRACT

Demographics in Malaysia and Viet Nam are evolving rapidly, potentially disrupting traditional family support to older people. We estimate a set of Poisson random effects models with panel data from the Malaysia Ageing and Retirement Survey and the Viet Nam Aging Survey to analyze how living arrangements, marital status, and support from children influence the mental and physical health of older people. In Malaysia, having living children plays an important protective role for both mental and physical health, while living with a son appears to have a protective effect for physical health. Results are similar for Viet Nam, except older women, who are at greater risk of mental and physical health problems, appear to enjoy a greater protective effect for their mental health from a child living nearby than do men. Our analysis underscores the importance of social safety nets for the health of senior citizens living alone.

_________________________

The authors thank Aiko Kikkawa, Minhaj Mahmud, Maki Nakajima, Donghyun Park, participants in the 2023 Asian Development Bank workshop on the State of Well-Beings of Older Persons in Asia, and participants in the 2024 Asian Development Bank Economists' Forum for their helpful suggestions. Guidance from Norma Mansor on the use of the Malaysia data and from Long Thanh Giang on the use of the Viet Nam data was greatly appreciated.

**I. Introduction**

The number of older Asians aged 65 and above is growing rapidly. By the year 2050, the share of Asia's older people in the total population will be 18%, exceeding the global average of 16% (Asian Development Bank [ADB] 2017). Although developing countries in the region still have relatively young populations, their demographic transitions are progressing rapidly, with a considerable shift in the age structure toward older people. This shift has prompted legislative reforms to build and reinforce the social safety net to better support older people, especially those living in poverty.

A growing body of research is examining how the well-being of older women and men is influenced by their living arrangements. Across countries, families are using their own unpaid labor, especially that of women, to provide care for older family members (Stark 2005). Unpaid care for older men is often provided by spouses, while for women it is provided by other family members, especially by daughters, as documented for the Republic of Korea (Yoon 2014) and the People's Republic of China (Chen et al. 2018). One outcome of these informal care arrangements is lower hospitalization rates for older people living with family members compared with the older people who live alone (González-González et al. 2011). In what has been called the feminization of later life, the population share has become increasingly female among older age groups in most countries, with women having higher life expectancies than men on average because of a combination of biological, social, and behavioral reasons (Kinsella 2000). Given women's longer life expectancies, in most countries relatively more women are widowed than men, leaving women more reliant on formal services outside of the home for their healthcare needs, including long-term care.

Despite their longer life expectancies, women are more likely than men to report health problems and to have worse self-reported health status (Atchessi et al. 2018; Madyaningrum, Chuang, and Chuang 2018; Case and Paxson 2005). Older women are more vulnerable to poverty given their relatively higher rates of widowhood as well as insufficient support from pensions and social security owing to their relatively shorter times in the labor force (Smeeding and Sandstrom 2005). In turn, poverty and income constraints contribute to problems in accessing healthcare. Older women have faced additional constraints, including relatively lower rates of health insurance coverage, less education, and lack of economic independence, that have limited their access to healthcare and their ability to pay for it (Brinda et al. 2015, Zhang et al. 2017). Gender discrimination and class bias against poor people may also play a role causing women to slip through the cracks in gaining access to services from network hospitals (Karpagam, Vasan, and Seethappa 2016).

We contribute to this literature by examining factors that are associated with the well-being of older people in Southeast Asia, with a focus on the role of family support and living arrangements. We study two countries that are experiencing pronounced demographic transitions and evolving challenges faced by the older members of their populations: Malaysia and Viet Nam. Our goal is to determine how support from children influences the mental and physical health of older people, and how these effects differ by ethnicity and gender. A focus on differences across ethnic groups reveals how people with varying ethnic backgrounds view the role of family in caring for aging parents. This approach is motivated by qualitative evidence that older Chinese in Malaysia are not supported by their children and are placed into nursing homes instead (Tey et al. 2016).

Moreover, the analysis will also show how discrimination by ethnicity—similar to the case of gender bias—may be reflected in disparities in the well-being of older people, to the extent that some ethic groups in these countries have experienced bias and marginalization in the labor market.

Cultural nuances and different kinship systems can play a major role in influencing how older parents respond to care by their children, as evidenced by previous research indicating that the psychological well-being of older people in Viet Nam is positively associated with living with a married son but not with a daughter, while in Thailand, co-residence with a child has beneficial effects for the psychological well-being of older people regardless of the gender of the child (Teerawichitchainan, Pothisiri, and Long 2015). Our objective is to build on this research along several dimensions: we look at three measures of well-being, we use more recent data for Viet Nam and new panel data for Malaysia, and we explore how the relationship between living arrangements and the well-being of older people varies by gender and ethnicity.

## II. Background

### A. The Demographic Transition

Malaysia is still in the early stages of the demographic transition while Viet Nam is (somewhat surprisingly, considering its still relatively low income) at a more advanced stage. In fact, Viet Nam is one of countries of the Association of Southeast Asian Nations that is widely cited as being at risk of getting old before getting rich (Huong 2017). Therefore, the demographic profiles of Malaysia and Viet Nam are quite different in terms of the speed of population aging. Figure 1 depicts population pyramids for the years 1990 and 2021 in which the bars indicate the percentage of the total population that is male or female and belongs to a particular age group. Both countries have clearly undergone pronounced demographic transitions since 1990, with a large shift in the population distribution from children to adults. Consistent with other middle-income countries in the region, the population is now concentrated among age groups considered to be young and working-age adults, and Viet Nam has started to see that shift in the distribution of the population toward older people. Strikingly, gender imbalances are progressively skewed toward women in older age groups, especially in Viet Nam.

*Insert Figure 1 here*

Looking at the data from another angle shows that most of the older people are women. Figure 2 shows the male–female sex ratio by 5-year cohorts for both countries; that is, the ratio of the number of males in each 5-year age group to the number of females, expressed as a percentage. Numbers close to 100 indicate that the shares of males and females in an age group are roughly the same. These ratios exhibit a marked and sharp drop for the older population groups, especially after the age of 50, and especially in Viet Nam. Overall, among older people, women make up the majority share for all age groups, and the difference is particularly stark for those aged 70 and above.

*Insert Figure 2 here*

## B. Aging and Family Support in Malaysia

A small but growing body of research has examined the essential role of social support for both physical and mental health outcomes for older people, especially concerning depression and physical inactivity (Sazlina et al. 2012, Marthammuthu et al.

2021, Sahril et al. 2023, Yahaya et al. 2013). For example, Kooshair et al. (2014) find that: (1) fewer social support functions are available for older women than for older men, (2) gender moderates the effect on the relationship between positive social interaction and life satisfaction, and (3) social support from children and family members has a significant positive effect on life satisfaction. Cultural traditions also play a role, in that relationships, gender roles, social norms, and group cohesions usually matter more for people's behaviors than their own beliefs (Kooshair et al. 2014). A study of more than 2,000 Malaysian older people living across the country found that the quality of social networks may matter more than living arrangements (Hamid et al. 2021). Women in Malaysia are often found to be at greater risk for several negative outcomes related to well-being, including frailty, anemia, dementia, weight problems, and security or safety at home (Badrasawi, Shahar, and Singh 2017; Yusof et al. 2018; Hamid et al. 2010; Chen, Ngoh, and Harith 2012; Suriawati et al. 2020).

In Malaysia, discrepancies in health by ethnicity can be pronounced. Conditions like arthritis, high blood pressure, diabetes, asthma, and perceived health status vary considerably across ethnic groups. After controlling for socioeconomic and health and lifestyle factors, Teh, Tey, and Ng (2014a) found that Chinese older people are least likely to report poor health, whereas Indians and indigenous peoples are more likely to do so. Further, older Malays have a lower likelihood than older Chinese and Indian of undergoing check-ups due to the influence of cultural beliefs (Cheah and Meltzer 2020). These ethnic disparities in seeking medical check-ups appear only in lower-income groups, not middle-income or high-income groups. Closely related, Omar (2003) found disparities by ethnicity in social support in the city of Petaling Jaya: older men, especially Muslim and Chinese men, were more mobile, whereas older women were more housebound, experienced greater loneliness, and were more likely to suffer from depression. Issues related to the migration of adult children compound these differences across ethnic groups. In particular, semi-structured interviews of older Chinese and Malay adults indicate that assistance from their emigrant adult children tended to be mostly informational and financial, often substituting for a lack of more tangible support (Evans et al., 2017). Although the adults in this purposive sample lived alone by choice, there was quite some ethnic variation in types of support. Older Malays received more support from nearby adult children and relatives, whereas older Chinese adults were less likely to have adult children living locally and more likely to emphasize support from friends and neighbors instead.

## C. Aging and Family Support in Viet Nam

Viet Nam exhibits substantial regional differences in the patriarchal family system. Older people are much more likely to reside with a married son than daughter, and this tendency is more pronounced in the northern regions of Viet Nam than in southern and central areas (Friedman et al. 2003). In contrast, Viet Nam exhibits less variation by gender among older people when it comes to intergenerational transfers, household wealth, and self-perceptions of economic satisfaction (Friedman et al. 2003). Self-reported health status is a different matter, as older Vietnamese women are more likely than men to report poor self-rated health, with factors like social participation playing a confounding role (Le, Quashie, and Prachuabmoh 2019). More specifically, additional predictors of health for older women include living alone, number of children living nearby, and average frequency of talking on the phone with children, while satisfaction with respect from the community, financial support from children, and information support are strongly associated with older men's health (Giang, Nguyen, and Nguyen 2020). However, while intergenerational co-residence was found to increase the psychological well-being of older people in Viet Nam, this effect was more important for older men than for older women (Yamada and Teerawichitchainan 2015).

Women comprise about three-quarters of older people in Viet Nam who live alone, and they are more vulnerable to health problems, especially anxiety and depression. (Vo et al. 2021, Pham et al. 2018). About 30%–60% of older women in rural Viet Nam self-reported moderate health problems related to mobility, self-care, usual activities, pain or discomfort, and anxiety or depression (Hoi, Chuc, and Lindholm 2010). Van Minh et al.'s (2010) study of the Bavi District, a rural community west of Ha Noi, affirm these general findings with evidence that a higher proportion of older women reported both poor health status and poor quality of living compared to older men.

In terms of healthcare-seeking behavior among older people, previous research suggests that Viet Nam's long history of socialized medicine helps to explain why the country sees a relatively high probability that individuals seek healthcare services from professional providers compared to some regional neighbors (Rodgers and Zveglich 2021). However, in Viet Nam, being a woman is negatively associated with healthcare-seeking behavior. A possible reason is that women face relatively more stigma and discrimination than men in seeking treatment for some communicable diseases in Viet Nam (Govender and Penn-Kekana 2008, Van Minh et al. 2018).

### III. Data and Methodology

The study uses panel data from the Malaysia Ageing and Retirement Survey for 2018–2019 and 2021–2022 and the Viet Nam Aging Survey for 2019 and 2022. The Malaysia Ageing and Retirement Survey data, collected by the Social Wellbeing Research Centre, is national longitudinal data based on in-person interviews with adults aged 40 years and above. Wave 1 was completed in 2019 with 5,613 respondents, and Wave 2 was completed in 2022 with 4,821 respondents (of which 75% consisted of panel respondents who participated for the second time). The Viet Nam's Institute of Social and Medical Studies collected the Viet Nam Aging Survey data through a home-based survey to evaluate social health insurance for aging adults, with 4,333 respondents aged 50 and above in 2019. The 2022 wave only included those aged 60 and above, and it had 3,183 respondents (of which 45% consisted of panel respondents who participated in the 2019 wave). Both the Malaysia and Viet Nam data sets have detailed information on socioeconomic and demographic characteristics, family relationships and support, health outcomes, and healthcare utilization. For both countries, we run regressions for the full sample as well as a sample of older people, defined to be those individuals aged 60 and above (consistent with the United Nations' definition of older people).

We model the determinants of well-being in old age using Poisson random effects regressions, following the approach of Díaz-Venegas, Sáenz, and Wong (2017). Poisson regressions, which are used to analyze count data such as the number of depressive symptoms reported by a respondent, can be used to explore how various factors can predict the likelihood or frequency of an event occurring. We focus on two indicators: the number of self-reported depressive symptoms and the number of self-reported chronic conditions.

For the Malaysia survey, respondents were asked about the frequency of 18 possible depressive symptoms they may have experienced in the previous 6 months. The symptoms include negative experiences (such as sadness, losing interest in most things, and thoughts of death) and positive experiences (such as feeling cheerful, satisfaction with life, and having people to turn to). Responses for all 18 questions were coded as a Likert scale, ranging from 1=never to 5=always. To construct our indicator of the number of depressive symptoms, we counted the number of negative experiences for which "always" or "often" was the response plus the number of positive experiences for which "never" or "rarely" was the response. For the number of chronic physical conditions, respondents were asked if they had ever been diagnosed by a doctor for 19 possible illnesses, including cancer, stroke, heart disease, and diabetes. For this indicator, we simply counted the number of conditions for which the response was yes. Variable construction using the Viet Nam data was similar, with some notable differences. For the 15 depressive symptoms, allowable responses were yes or no rather than a Likert scale, and the reference period was the previous 7 days. The variable for the number of chronic physical conditions, based on 15 listed conditions, was constructed in the same way as the Malaysian data.

One of the key independent variables is the number of living children, with the a priori expectation that the number of children is beneficial for the well-being of older people. In both countries, adult children are usually considered the traditional care providers and the most important source of support for older people. Because this association may be nuanced by living arrangements and the gender of the children, our regressions include control variables for these indicators. In particular, we have a set of dummy variables for living arrangements: living with a spouse only, living with other family members, and living alone (the excluded variable). We also include controls for gender, ethnicity, educational attainment, age, marital status, geographical region, and survey wave. In Malaysia, the ethnic groups are Malay (majority), Chinese, Indian, other Bumiputra, and other groups. In Viet Nam, the ethnic groups are Kinh (majority) and other groups. For Malaysia only, we also include a set of dummy variables for the location of the nearest living son: living with the respondent, living near the respondent, and living elsewhere (the excluded category); a similar set of variables is included for the nearest living daughter. For Viet Nam, a set of dummy variables is included for the location of the nearest living child, though the survey did not have questions that allowed us to create such variables by gender (son or daughter).

Because we are interested in how the effect of family support differs by ethnic group and by gender, for some specifications, we construct sub-samples based on gender and ethnicity (constructed as ethnic majority and an aggregation of the ethnic minorities) and run separate regressions for those sub-samples. The results for the gender sub-samples are reported in the main tables, and the results for the ethnic sub-samples are available upon request. All statistical analyses are weighted to the population using the sampling weights from the respective data sets. Standard errors are corrected for clustering at the household level. The online appendix[1] provides detailed variable definitions (Appendix Table 1) and sample means (Appendix Tables 2 and 3).

---

[1] The Appendix is available at http://dx.doi.org/10.22617/WPS240325-2.

Figures 3–6 compare several key variables between the two countries. As shown in Figure 3, 57% of older people in Viet Nam report having at least two chronic physical health conditions, considerably higher than Malaysia (32%); only 18% in Viet Nam were not diagnosed with any chronic conditions compared to 46% in Malaysia. Older people in Viet Nam are also more likely to report having depressive symptoms: 57% of older people in Viet Nam reported experiencing at least three depressive symptoms compared to 18% in Malaysia. Also of note are considerable gender differences in marital status in both countries: older men are considerably more likely to still be married compared to older women. In contrast, older women are much more likely to be widowed in both countries, especially in Viet Nam—presumably as a long-term repercussion of the loss of men's lives during the Viet Nam war (Figure 4). In both countries, the vast majority of older people live with other family members (85% in Malaysia and 72% in Viet Nam) (Figure 5). Gender differences in living arrangements are less pronounced in Malaysia than they are in Viet Nam, where 67% of older men live with other family members compared with 74% of older women. Also in both countries, older women are more likely than older men to live by themselves, while the opposite is true for living only with one's spouse. A final point of interest in the descriptive statistics is the location of the nearest living child. A large proportion of older people in both countries live with their children: in Malaysia, 77% of older people live with at least one child, and in Viet Nam that share is 65% (Figure 6). In Malaysia, more older people live with a son than with a daughter, and this is true for both men and women.

*Insert Figures 3-6 here*

**IV. Estimation Results**

**A. Malaysia**

Table 1 shows the results for the factors that are associated with the number of depressive conditions in Malaysia. The table reports Poisson coefficients, each of which is interpreted as the expected change, on a log scale, the count of depressive conditions for a one-unit change in the predictor variable (holding the other predictor variables constant).

*Insert Table 1 here*

Looking first at the results for the overall sample in column 1 and the two age-group subsamples (aged 40–59 and aged 60+) in columns 2–3, we see that increasing age is associated with a somewhat higher expected number of depressive symptoms, but gender has no statistically significant association with depressive symptoms. Columns 2 and 3 show that the positive association between age in years and depressive conditions is coming entirely from the older people aged 60 and above. Among the ethnic groups in the overall sample, compared to the other Bumiputra category, being Indian is associated with more depressive symptoms, and this is driven mostly by the those aged 40–59. In contrast, being Chinese is associated with fewer depressive conditions for both age groups and for the overall sample. This result is consistent with previous findings in the literature that among older people in Malaysia, Chinese individuals are less likely to self-report poor health compared to Malays, Indian, and Indigenous People, and this result holds across income groups (Teh, Tey, and Ng 2014a; Chan et al. 2015; Khan and Flynn 2016). A possible reason is that older Chinese adults exhibit more health-promoting behaviors than older Malay and Indian adults (Mohd et al. 2022).

Our key independent variables for the number of children and living arrangements yield some meaningful results. As shown in column 1, compared to adults with no living children, adults who have at least one living child have a substantially lower expected number of depressive symptoms, and this is particularly true for adults who have three or four living children. This protective effect of having living children is only true for older people aged 60 and above. Closely related, compared to adults who live alone, adults who live with their spouse or with other family members have a lower expected number of depressive symptoms, and this is especially true for adults who live only with their spouse. Again, columns 2 and 3 show that these living arrangement variables only have statistically significant coefficients for older people aged 60 and above. Although it does not appear to matter if the adult is living with a son or daughter in the same home, it does matter for mental health to have a daughter living nearby, but in an unexpected way: living near one's daughter is associated with a higher number of depressive symptoms. This result is driven by older people aged 60 and above. Closely related to living arrangements is marital status: adults who are separated or divorced have more depressive symptoms compared to people who are married in both age-group samples. Being widowed is also associated with more depressive symptoms, but this result is explained mostly by adults aged 40–59. Having never been married also appears to have an adverse mental health effect for adults aged 40–59 but a protective effect for older people aged 60 and above.

Results for the other independent variables in columns 1–3 are as expected. More education is generally associated with fewer depressive conditions compared to having no education at all, and the magnitude of this relationship is larger the higher the educational attainment. In addition, living in an urban area is associated with a lower number of expected depressive conditions, although the opposite is true for living in Peninsular Malaysia where most of Malaysia's population lives.

Columns 4–6 show how these results for the two age-group samples differ by gender of the respondent. Of note, the increase in depressive conditions associated with age is particularly strong for women aged 60 and beyond, while the protective effect of being Chinese holds for both men and women (except that the coefficient is not statistically significant for men aged 40–59). The results for the number of living children and living arrangements are also striking. The protective effects of having living children compared to having no children for older people hold for both men and women, as do the beneficial effect of living with one's spouse only. However, only older men seem to experience the adverse effect for mental health from living near a daughter.

Results for the number of chronic health conditions for Malaysia are in Table 2. Looking first at the overall results and the two age-group subsamples in columns 1–3, we see that being a woman is associated with a higher expected number of chronic health conditions for older people aged 60 and above, but not for younger adults. In contrast, an additional year of age is linked with more chronic conditions for adults aged 40–59, but not for older people. Among the ethnic groups, the largest effect is found for being Indian, which is associated with a fairly large increase in the expected number of chronic conditions among adults in their 40s and 50s, but not older people. Ethnic Malays in their 40s and 50s also have more chronic conditions, while the opposite is true for the "other" ethnic category. Again, these results for the different ethnic groups are largely in line with previous studies on self-reported health among Malaysia's different ethnic groups (Teh, Tey, and Ng 2014a, Chan et al. 2015, Khan and Flynn 2016).

*Insert Table 2 here*

Among the results for the number of children and living arrangements, having at least one child is positively linked with the number of chronic conditions, and this effect is coming entirely from older people aged 60 and above. The general living arrangements do not appear to play much of a role, but living with or near one's children does. Living near one's daughter is associated with a higher number of chronic conditions, especially for adults aged 40–59, while living with one's son is associated with fewer chronic conditions, especially for older people aged 60 and above. Because these estimates are associations and not causal effects, it could be that daughters are likely to live close to their parents if their parents are in poor health with multiple chronic conditions, but the daughters' own childcare arrangements prevent them from living with their parents. For older people, and especially older men, living with a son has a clear protective effect for physical health. Being widowed is also associated with having a higher number of chronic conditions. As for the other control variables, the most notable finding is the lack of a consistent protective effect from educational attainment, which is counter to the results we found for depressive conditions.

As for gender differences in Columns 4–6 of Table 2, there are several key findings. First, only women in their 40s and 50s, not men, report a positive association between being Malay and having more chronic health conditions. In contrast, the positive association between having at least one child and having more chronic health conditions among older people aged 60 and above holds more for men than it does for women. A similar conclusion applies to living near one's daughter: among adults aged 40–59, the positive association between living near one's daughter and number of chronic conditions is considerably larger for men than for women. Being widowed also has a stronger positive association with chronic conditions for men than women among older people aged 60 and above.

## B. Viet Nam

Results for the predictors associated with the number of depressive conditions in Viet Nam are in Table 3. Overall, results for adults aged 60 and above are very similar to results for the overall sample, so compared to Malaysia, there is far less nuance by age in Viet Nam, which may be partly accounted for by differences in minimum ages in the two countries' samples. Overall, and among older people above the age of 60, being a woman is associated with having a higher number of depressive symptoms, and the magnitude of this association is fairly large compared to many other estimates. However, being part of the ethnic majority (Kinh) has a protective effect on mental health. Among the estimates for the number of children and living conditions, having at least one child is associated with a lower number of depressive symptoms for both men and women, and the same is true for living with their spouse or living with other family members, as compared to living alone. Compared to not living close to any children, living with a child or living near a child also appears to have a protective effect on mental health for all adults and for older people.

*Insert Table 3 here*

There are more variations by age and gender for marital status. In particular, being separated or divorced is associated with a lower expected number of depressive symptoms for the overall sample but a higher number of depressive symptoms for older people aged 60 and above. Most of that effect is coming from older women. Also contributing to having more depressive symptoms for women but not for men is being widowed. In direct contrast, never having been married is associated with more depressive symptoms for men and fewer depressive symptoms for women. Also of note is the protective effect of having higher educational attainment on the mental health of both older men and older women; we saw a similar result for Malaysia. Moreover, the protective effect of living with or near a child appears to only hold for women, as men have more depressive symptoms if they are living with or near a child.

Finally, Table 4 reports the results for the number of chronic health conditions in Viet Nam. As we saw with the results for depression, being a woman is associated with a higher expected number of chronic health conditions than men. Age and being ethnic Kinh also have a positive association with the number of chronic health conditions. In contrast, living with or near a child is strongly associated with fewer chronic health conditions for all adults and for older people.

*Insert Table 4 here*

The results for support from family members show more differentiation by age and gender. In the overall sample, having at least one child is associated with a higher number of chronic conditions, and this result is driven primarily by men. Among adults in their 60s and above, the relationship is less clear-cut and varies by the number of children and by the gender of the respondent. So, for example, having at least one child is associated with a meaningful increase in the expected number of chronic conditions for older men, and this risk varies with the number of children, but for older women, the relationship is smaller and in one case (having 3–4 children) it even becomes negative. As for living arrangements, for the most part, there is a protective effect from living with their spouse only or with other family members. This holds for everyone except for older women aged 60 and above, for whom living with their spouse is associated with more chronic health conditions. Moreover, the apparent protective effect for physical health from living with or near a child is larger for men than women.

Men and women differ considerably in how marital status is associated with the number of chronic conditions. While widowhood and being separated or divorced are associated with fewer chronic health conditions among older men in their 60s and beyond, the association becomes positive for older women. However, never having been married appears to be good for one's physical health for both older men and older women. The final noteworthy result is that the protective effect we saw in the case of mental health does not appear to hold for chronic physical conditions. Having secondary or tertiary schooling is associated with more chronic conditions for the overall sample and the sample of older people, and it holds for both men and women except for the case of older women having tertiary education.

## V. Conclusion

This study has explored factors that are associated with health outcomes of older people in Malaysia and Viet Nam. We used recent health surveys to estimate Poisson random effects regressions to model the determinants of well-being in old age, with a focus on mental health and physical health. Our main findings for Malaysia point to a greater incidence of self-reported chronic health conditions among older women relative to older men. Consistent with earlier studies, we also find substantial differences across ethnic groups in the likelihood of reporting both depressive symptoms and chronic health conditions, with Indian adults being more likely to report mental and physical health issues compared to Chinese adults. Having living children plays an important protective role for both mental and physical health compared to having no children at all, especially for older people aged 60 and above. Living with a son also appears to have a protective effect on physical health. As expected, living alone and being widowed or separated or divorced are all associated with more depressive symptoms. The results for Viet Nam are similar, except that older women are at greater risk of both mental health and physical health problems compared to men, and women in Viet Nam enjoy more of a protective effect for their mental health from living with or near a child than do men. On balance then, family support is critical for the mental and physical health of older people in both countries, with the implication that senior citizens who are living alone need extra support through the social safety net.

In Malaysia, the government has taken several steps to promote the health and well-being of older people. Notably, Malaysia is one of just a few Asian countries that has achieved universal health coverage, meaning that people receive needed healthcare services without experiencing financial hardship (Kowal, Ng, and Hoang 2024). Another key measure in place includes the National Policy on Aging, which covers various aspects of older people's well-being, including healthcare, social support, and financial security. The government has been working to increase the accessibility and affordability of healthcare services, as well as providing specialized services for age-related conditions. The government has also conducted health education and promotion campaigns targeting older people to raise awareness about healthy living, disease prevention, and the importance of regular health check-ups. In terms of financing healthcare, the government provides financial support to older people through programs like the Social Welfare Department's financial aid schemes, which aim to assist those with low incomes. Such efforts have not only come from the top down, as initiatives have been implemented to provide community-based care for older people, which allows them to receive healthcare services and support in their own communities, thus helping to reduce the need for institutionalized care. In addition, various social support programs and initiatives have been launched to address the social and emotional well-being of older people, including social clubs, recreational activities, and counseling services.

More recent efforts include implementing and enforcing regulations on elder care services in Malaysia to ensure that older people receive appropriate care and support. These various initiatives reflect the government's recognition of the aging population and the need to focus on care. In early 2023, the government proposed a comprehensive policy solution to address its aging population. One of the most striking provisions of the Senior Citizens' Bill is its financial penalties for people placing their older parents into formal care institutions, highlighting the continued role of filial piety customs in Malaysian culture by attempting to enforce a statutory duty that legally binds adult children to caring for their older parents (Hui 2023). However, critics point to the 2023 and 2024 national budgets as evidence that the Government of Malaysia lacks a strong focus on responding to the needs of its aging population (New Straits Times 2023, Thomas 2023). The Malaysian Coalition on Aging has continued to urge the government to do more to support the healthy and active aging of older people, emphasizing limited savings of older people and advocating for a basic universal pension scheme. Hui (2023) also proposes greater investment in Malaysia's care economy and care infrastructure to respond to the vast majority of older people's preferences to age in their own homes.

In Viet Nam, the older people share of the total population is projected to increase from 8.1% in 1999 to almost 20% by 2035 (United Nations Population Fund [UNFPA] 2019). The Government of Viet Nam has recognized the growing size of its older population and has already implemented several policy measures to address their needs. These efforts include an Ordinance on Elderly People, passed in 2000, that contained provisions for the support for and care of older people. In 2009, this Ordinance was replaced by a broader Law on the Elderly 2009, which guaranteed the rights of older people. It was followed 3 years later by the National Action Program on the Viet Nam Elderly, which contained specific social targets, including health care and the promotion of "active aging" (UNFPA 2019). In 2014, the government instituted a revised Health Insurance Law that removed barriers to coverage faced by poor people (Thuong 2020), and it has since adopted additional resolutions to address the needs of the aging population further.

Viet Nam's fast-aging population has created new policy challenges, especially in terms of welfare administration. Structural problems with its national welfare system, particularly through the separation of the pension and social welfare system, have left many older people behind. Further, low levels of participation in Viet Nam's social insurance program (only 38% as of June 2023), and continued withdrawals from the program have threatened pension security, placing older people at greater risk (Giang 2023, Nguyen 2023, Viet Nam News 2023b). The government has tried to address these problems. In 2019, it enacted a policy to phase-in increases in the retirement age, from 60 to 62 for men by 2028 and from 55 to 60 for women by 2035 (Webster et al. 2019). Moreover, in August 2023 the government passed a resolution to study and propose a social allowance and pension program for those aged 75–80, along with general raises in social allowances for the older population (Viet Nam News 2023a). If stigma and discrimination in seeking treatment are contributing to Viet Nam's sizeable male–female gap in healthcare-seeking behavior, then programs and policies that adjust these types of attitudes and gender norms will go a long way to eliminate health inequities in Viet Nam.

Like Malaysia and Viet Nam, many countries in the region have implemented reforms that focus on the needs of their aging populations, but progress has remained uneven, depending on the size of public sector health budgets (Mahal and McPake 2017). Governments in the region will need to set out plans to better prepare their populations for active and healthy old-age living. The strategic challenge for policymakers across developing Asia is to equip current and future cohorts of older Asians with greater access to quality health, education, and other services, leading them to more fulfilling lives in their older age. From the society's perspective, success in this endeavor will determine whether the growing number of older Asians can become an asset rather than a burden.

**FIGURES AND TABLES**

## Figure 1: Population Pyramids for Malaysia and Viet Nam, 1990 and 2021

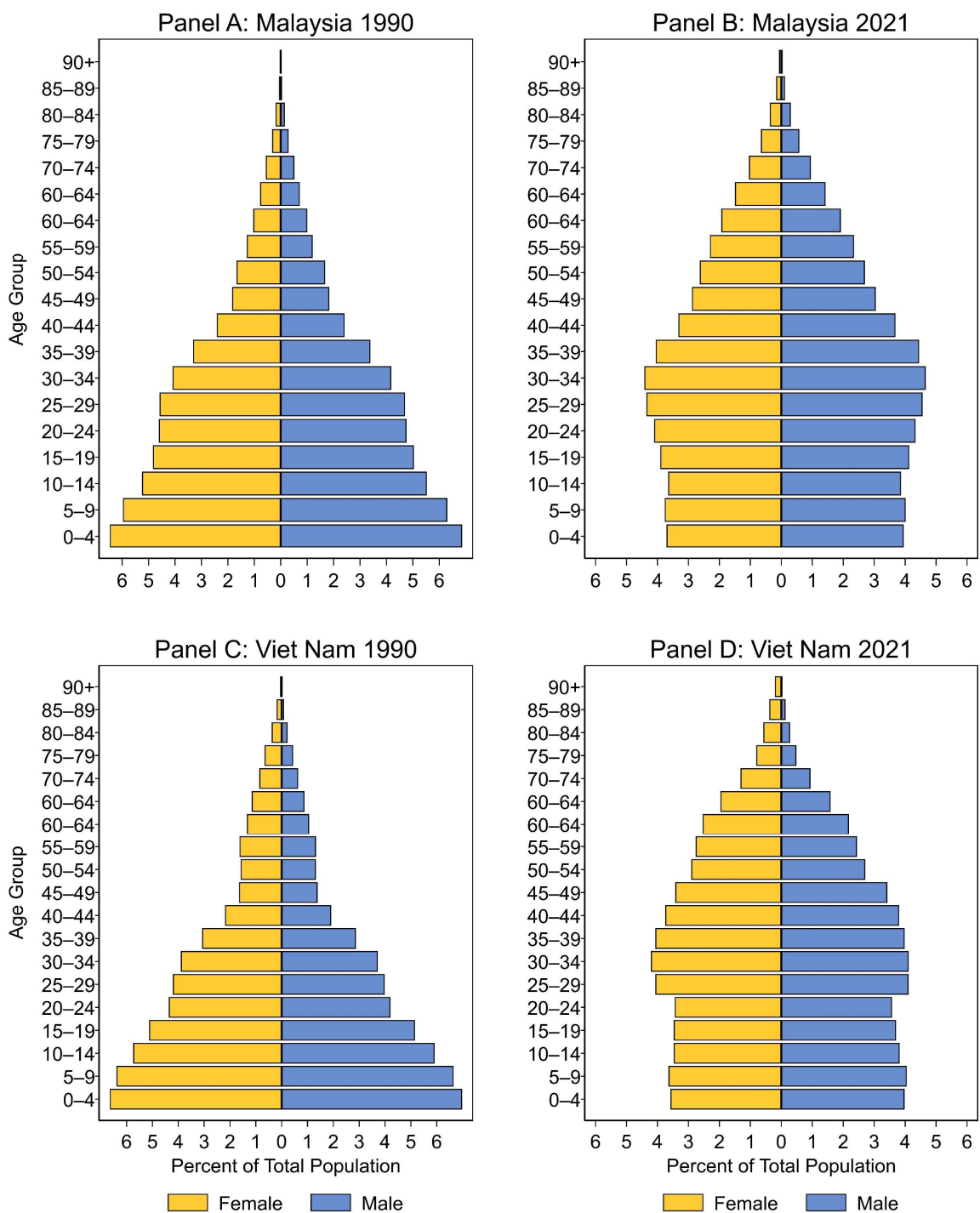

Source: Constructed with data from United Nations Department of Economic and Social Affairs (2022).

**Figure 2. Population Sex Ratios by Country, 2021**

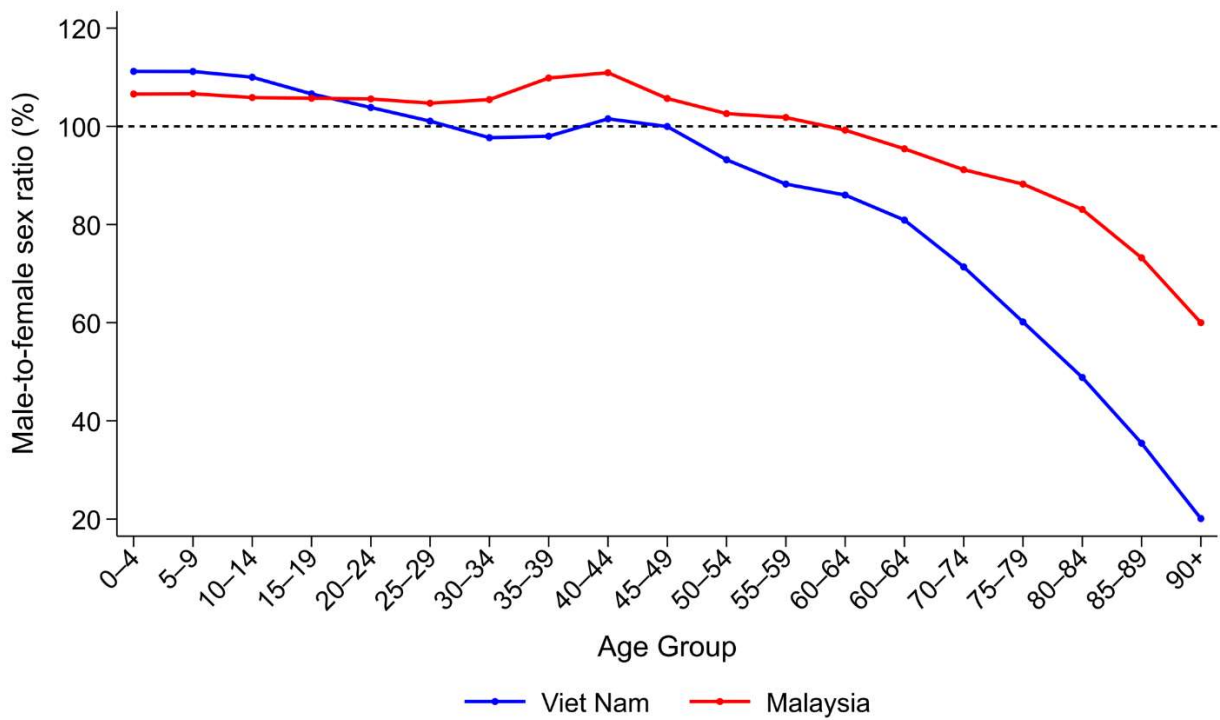

Source: Constructed with data from United Nations Department of Economic and Social Affairs (2022).

**Figure 3. Chronic Physical Conditions and Depressive Symptoms by Country**

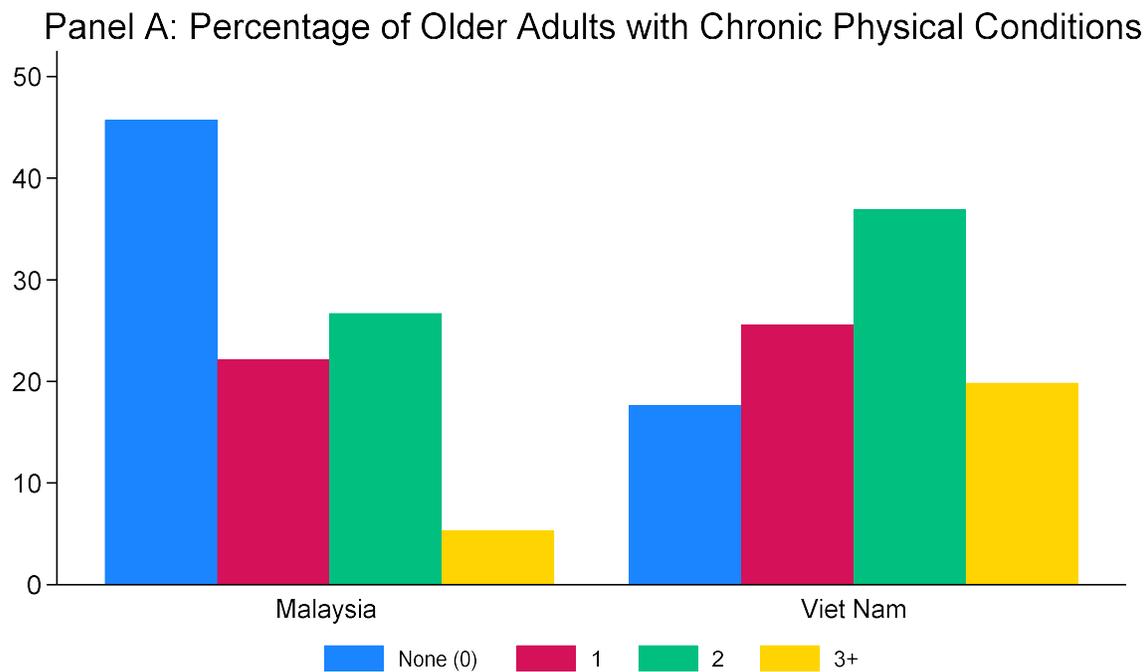

Panel A: Percentage of Older Adults with Chronic Physical Conditions

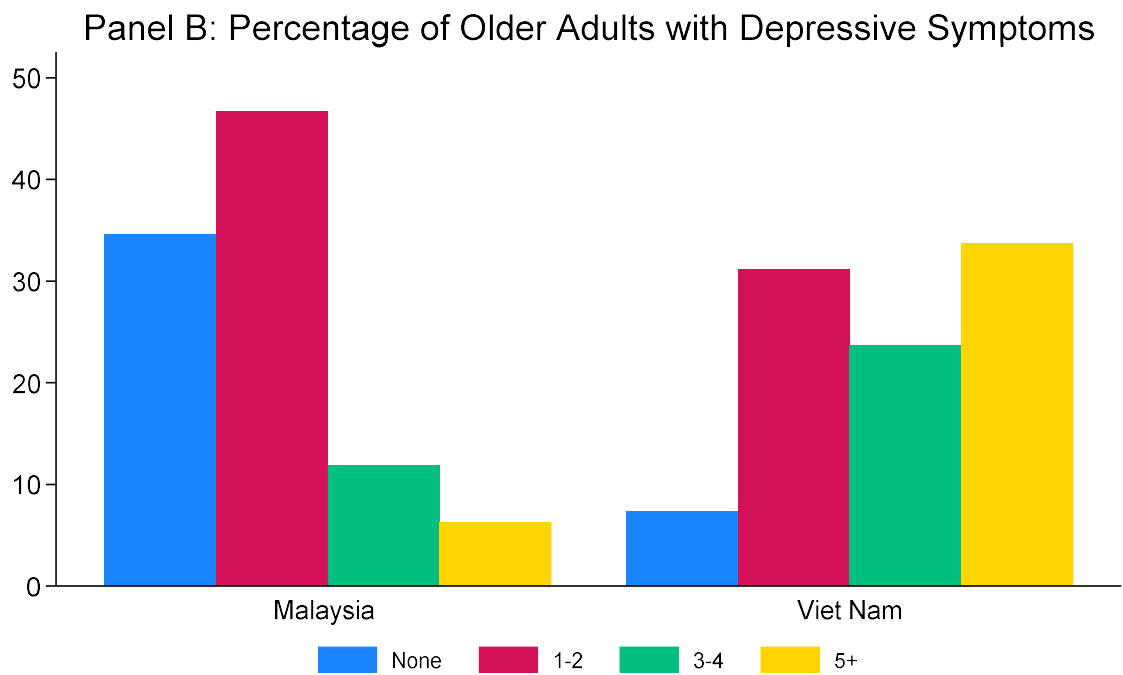

Panel B: Percentage of Older Adults with Depressive Symptoms

Source: Authors' calculations using Malaysia Ageing and Retirement Survey data and Viet Nam Aging Survey data.

**Figure 4. Marital Status by Country**

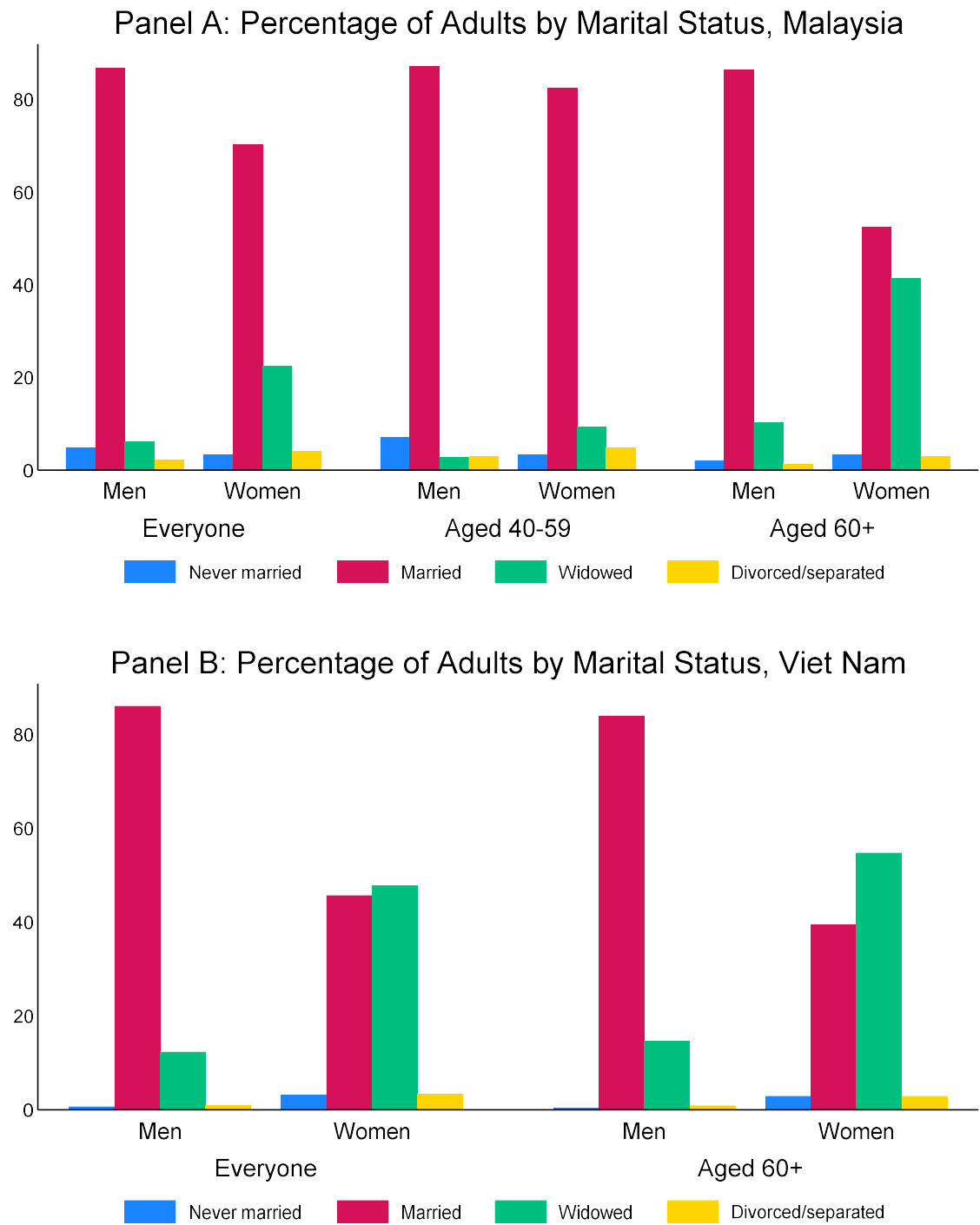

Source: Authors' calculations using Malaysia Ageing and Retirement Survey data and Viet Nam Aging Survey data.

**Figure 5. Living Arrangements by Country**

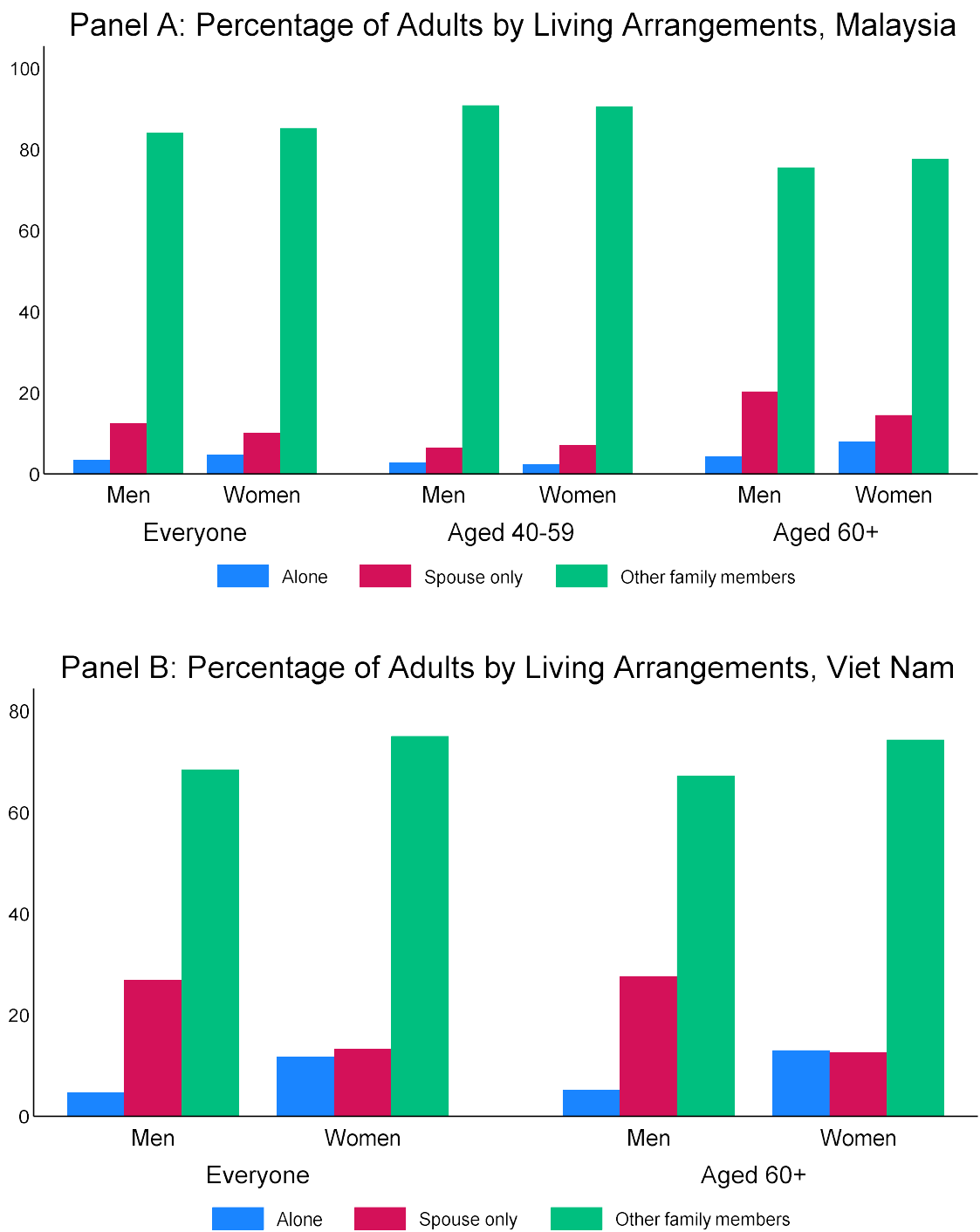

Panel A: Percentage of Adults by Living Arrangements, Malaysia

Panel B: Percentage of Adults by Living Arrangements, Viet Nam

Source: Authors' calculations using Malaysia Ageing and Retirement Survey data and Viet Nam Aging Survey data.

**Figure 6. Location of Nearest Living Child by Country**

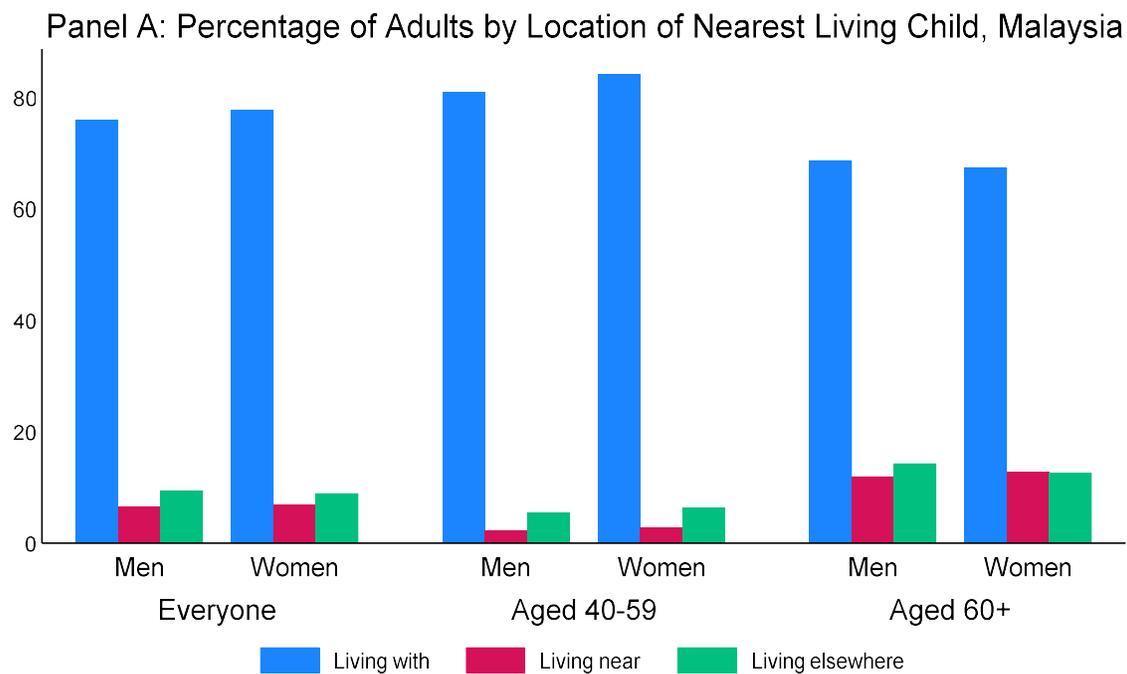

Panel A: Percentage of Adults by Location of Nearest Living Child, Malaysia

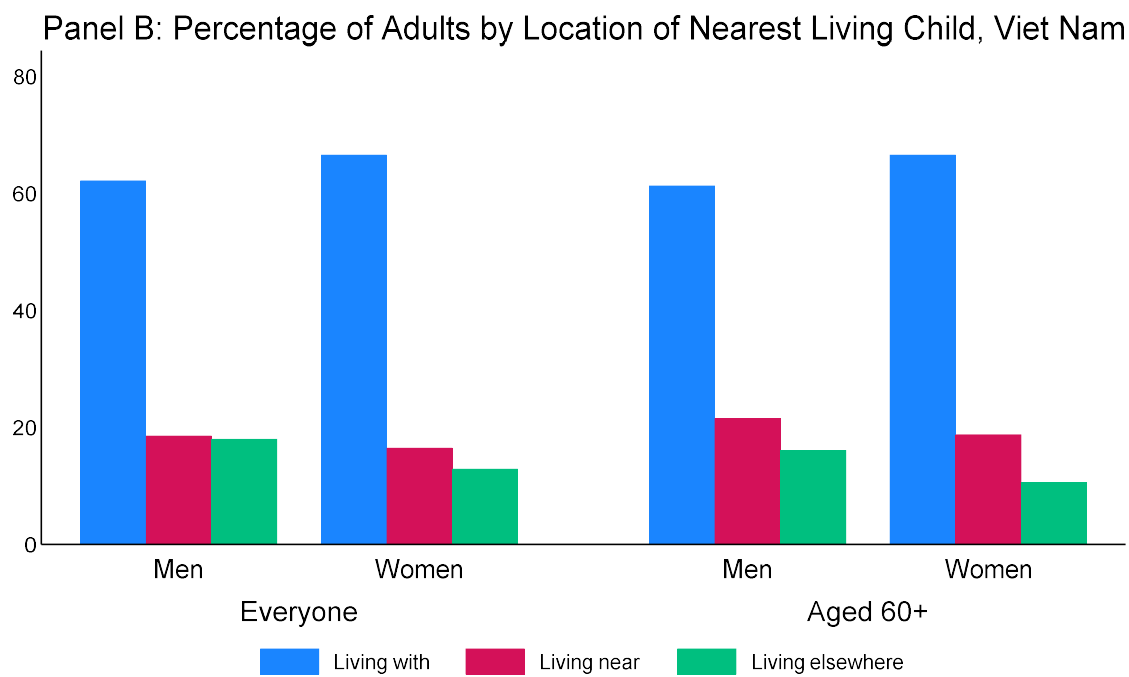

Panel B: Percentage of Adults by Location of Nearest Living Child, Viet Nam

Source: Authors' calculations using Malaysia Ageing and Retirement Survey data and Viet Nam Aging Survey data.

**Table 1. Poisson Random Effects Estimates for the Number of Depressive Symptoms, Malaysia**

| Variable/Statistic | All | | Aged 40–59 | | Aged 60+ | | Men Aged 40–59 | | Men Aged 60+ | | Women Aged 40–59 | | Women Aged 60+ | |
|---|---|---|---|---|---|---|---|---|---|---|---|---|---|---|
| Woman | -0.023 | | -0.049 | | 0.030 | | | | | | | | | |
| | (0.028) | | (0.039) | | (0.041) | | | | | | | | | |
| Age (years) | 0.003 | ** | 0.003 | | 0.016 | *** | 0.001 | | 0.010 | ** | 0.009 | * | 0.021 | *** |
| | (0.002) | | (0.004) | | (0.003) | | (0.006) | | (0.004) | | (0.005) | | (0.004) | |
| Ethnicity (reference: other Bumiputra) | | | | | | | | | | | | | | |
| Malay | -0.076 | | -0.103 | | -0.034 | | 0.014 | | -0.130 | | -0.209 | | 0.039 | |
| | (0.074) | | (0.100) | | (0.108) | | (0.156) | | (0.166) | | (0.133) | | (0.141) | |
| Chinese | -0.342 | *** | -0.372 | *** | -0.339 | *** | -0.120 | | -0.359 | ** | -0.599 | *** | -0.342 | ** |
| | (0.072) | | (0.099) | | (0.103) | | (0.153) | | (0.158) | | (0.134) | | (0.137) | |
| Indian | 0.185 | ** | 0.223 | * | 0.141 | | 0.366 | * | 0.025 | | 0.104 | | 0.233 | |
| | (0.090) | | (0.124) | | (0.128) | | (0.196) | | (0.203) | | (0.163) | | (0.165) | |
| Others | -0.071 | | -0.163 | * | 0.090 | | -0.187 | | -0.322 | * | -0.212 | * | 0.261 | ** |
| | (0.069) | | (0.091) | | (0.101) | | (0.139) | | (0.165) | | (0.124) | | (0.132) | |
| Education (reference: less than primary) | | | | | | | | | | | | | | |
| Primary | -0.141 | *** | -0.077 | | -0.140 | *** | -0.050 | | -0.205 | ** | -0.114 | | -0.104 | * |
| | (0.042) | | (0.082) | | (0.050) | | (0.148) | | (0.099) | | (0.099) | | (0.058) | |
| Secondary | -0.370 | *** | -0.333 | *** | -0.348 | *** | -0.223 | | -0.451 | *** | -0.401 | *** | -0.281 | *** |
| | (0.045) | | (0.080) | | (0.060) | | (0.139) | | (0.104) | | (0.096) | | (0.076) | |
| Tertiary | -0.644 | *** | -0.662 | *** | -0.519 | *** | -0.732 | *** | -0.584 | *** | -0.618 | *** | -0.533 | *** |
| | (0.081) | | (0.113) | | (0.126) | | (0.173) | | (0.161) | | (0.147) | | (0.206) | |
| Marital status (reference: married) | | | | | | | | | | | | | | |
| Never married | 0.003 | | 0.298 | ** | -0.496 | *** | 0.774 | *** | -0.375 | ** | -0.291 | * | -0.547 | *** |
| | (0.091) | | (0.127) | | (0.122) | | (0.173) | | (0.177) | | (0.160) | | (0.160) | |
| Widowed | 0.207 | *** | 0.407 | *** | 0.056 | | 0.317 | * | 0.181 | ** | 0.388 | *** | 0.002 | |
| | (0.040) | | (0.076) | | (0.047) | | (0.180) | | (0.079) | | (0.085) | | (0.059) | |
| Separated/ divorced | 0.447 | *** | 0.499 | *** | 0.384 | *** | 0.569 | *** | 0.613 | *** | 0.472 | *** | 0.259 | * |
| | (0.064) | | (0.079) | | (0.118) | | (0.117) | | (0.174) | | (0.101) | | (0.156) | |
| Urban | -0.110 | *** | -0.119 | *** | -0.097 | *** | -0.055 | | -0.107 | ** | -0.160 | *** | -0.086 | * |
| | (0.025) | | (0.036) | | (0.035) | | (0.054) | | (0.052) | | (0.048) | | (0.047) | |
| Peninsular Malaysia | 0.304 | *** | 0.357 | *** | 0.250 | *** | 0.259 | ** | 0.275 | *** | 0.421 | *** | 0.235 | *** |
| | (0.044) | | (0.063) | | (0.059) | | (0.102) | | (0.090) | | (0.080) | | (0.078) | |

**Table 1 (continued)**

| Variable/Statistic | All | | Aged 40–59 | | Aged 60+ | | Men Aged 40–59 | | Men Aged 60+ | | Women Aged 40–59 | | Women Aged 60+ | |
|---|---|---|---|---|---|---|---|---|---|---|---|---|---|---|
| Living children (reference: none) | | | | | | | | | | | | | | |
| 1–2 | -0.256 | *** | 0.064 | | -0.693 | *** | 0.352 | ** | -0.615 | *** | -0.131 | | -0.719 | *** |
| | (0.075) | | (0.108) | | (0.102) | | (0.166) | | (0.142) | | (0.127) | | (0.137) | |
| 3–4 | -0.322 | *** | -0.036 | | -0.686 | *** | 0.249 | | -0.550 | *** | -0.230 | * | -0.779 | *** |
| | (0.076) | | (0.113) | | (0.103) | | (0.178) | | (0.144) | | (0.135) | | (0.138) | |
| 5+ | -0.278 | *** | 0.122 | | -0.760 | *** | 0.486 | ** | -0.635 | *** | -0.135 | | -0.828 | *** |
| | (0.082) | | (0.124) | | (0.109) | | (0.192) | | (0.148) | | (0.149) | | (0.149) | |
| Living arrangements (reference: live alone) | | | | | | | | | | | | | | |
| Spouse only | -0.278 | *** | -0.164 | | -0.378 | *** | -0.148 | | -0.291 | ** | -0.159 | | -0.410 | *** |
| | (0.068) | | (0.121) | | (0.080) | | (0.164) | | (0.128) | | (0.175) | | (0.110) | |
| Other family members | -0.091 | * | -0.119 | | -0.111 | * | -0.117 | | -0.032 | | -0.109 | | -0.148 | * |
| | (0.054) | | (0.093) | | (0.067) | | (0.116) | | (0.114) | | (0.146) | | (0.084) | |
| Nearest living son (reference: living elsewhere) | | | | | | | | | | | | | | |
| Living with | -0.029 | | -0.078 | | 0.000 | | -0.153 | ** | -0.071 | | -0.022 | | 0.072 | |
| | (0.034) | | (0.049) | | (0.048) | | (0.077) | | (0.070) | | (0.063) | | (0.065) | |
| Leaving near | -0.076 | | -0.140 | | -0.025 | | -0.028 | | -0.214 | ** | -0.194 | | 0.104 | |
| | (0.053) | | (0.115) | | (0.060) | | (0.186) | | (0.093) | | (0.152) | | (0.079) | |
| Nearest living daughter (reference: living elsewhere) | | | | | | | | | | | | | | |
| Living with | -0.007 | | -0.025 | | 0.007 | | 0.067 | | 0.020 | | -0.073 | | 0.001 | |
| | (0.034) | | (0.049) | | (0.048) | | (0.075) | | (0.071) | | (0.064) | | (0.065) | |
| Leaving near | 0.097 | ** | 0.033 | | 0.149 | *** | 0.231 | | 0.271 | *** | -0.075 | | 0.070 | |
| | (0.047) | | (0.112) | | (0.054) | | (0.198) | | (0.078) | | (0.140) | | (0.073) | |
| Second survey wave | -0.137 | *** | -0.173 | *** | -0.084 | ** | -0.171 | *** | -0.052 | | -0.178 | *** | -0.110 | ** |
| | (0.025) | | (0.036) | | (0.036) | | (0.055) | | (0.053) | | (0.048) | | (0.048) | |
| Constant | 0.693 | *** | 0.432 | * | 0.234 | | 0.074 | | 0.585 | | 0.454 | | -0.081 | |
| | (0.147) | | (0.256) | | (0.259) | | (0.374) | | (0.395) | | (0.352) | | (0.334) | |
| No. of observations | 10,295 | | 5,961 | | 4,334 | | 2,553 | | 1,998 | | 3,408 | | 2,336 | |
| Model tests (p-values) | | | | | | | | | | | | | | |
| Model fit | 0.000 | | 0.000 | | 0.000 | | 0.000 | | 0.000 | | 0.000 | | 0.000 | |
| Random effects | 0.000 | | 0.000 | | 0.000 | | 0.000 | | 0.000 | | 0.000 | | 0.000 | |

*** = $p<0.01$, ** = $p<0.05$, and * = $p<0.10$.
Source: Authors' calculations using Malaysia Ageing and Retirement Survey data.

**Table 2. Poisson Random Effects Estimates for Number of Chronic Health Conditions, Malaysia**

| Variable/Statistic | All | | Aged 40–59 | | Aged 60+ | | Men Aged 40–59 | | Men Aged 60+ | | Women Aged 40–59 | | Women Aged 60+ | |
|---|---|---|---|---|---|---|---|---|---|---|---|---|---|---|
| Woman | 0.022 | | -0.029 | | 0.081 | ** | | | | | | | | |
| | (0.029) | | (0.047) | | (0.033) | | | | | | | | | |
| Age (years) | 0.032 | *** | 0.056 | *** | 0.003 | | 0.042 | *** | 0.008 | ** | 0.068 | *** | -0.002 | |
| | (0.002) | | (0.005) | | (0.003) | | (0.007) | | (0.004) | | (0.006) | | (0.003) | |
| Ethnicity (reference: other Bumiputra) | | | | | | | | | | | | | | |
|   Malay | 0.087 | | 0.255 | ** | -0.069 | | 0.081 | | -0.237 | * | 0.339 | ** | 0.071 | |
| | (0.074) | | (0.108) | | (0.094) | | (0.173) | | (0.144) | | (0.142) | | (0.125) | |
|   Chinese | -0.036 | | -0.030 | | -0.082 | | 0.051 | | -0.268 | ** | -0.158 | | 0.071 | |
| | (0.070) | | (0.105) | | (0.088) | | (0.165) | | (0.135) | | (0.138) | | (0.119) | |
|   Indian | 0.406 | *** | 0.722 | *** | 0.078 | | 0.710 | *** | -0.037 | | 0.693 | *** | 0.182 | |
| | (0.087) | | (0.129) | | (0.107) | | (0.203) | | (0.168) | | (0.168) | | (0.141) | |
|   Others | -0.349 | *** | -0.598 | *** | -0.179 | * | -0.677 | *** | -0.321 | ** | -0.552 | *** | -0.092 | |
| | (0.072) | | (0.107) | | (0.092) | | (0.168) | | (0.150) | | (0.137) | | (0.117) | |
| Education (reference: less than primary) | | | | | | | | | | | | | | |
|   Primary | 0.077 | * | -0.053 | | 0.042 | | -0.078 | | 0.137 | | 0.006 | | 0.018 | |
| | (0.044) | | (0.098) | | (0.046) | | (0.218) | | (0.100) | | (0.106) | | (0.052) | |
|   Secondary | -0.007 | | -0.216 | ** | 0.025 | | -0.037 | | 0.195 | * | -0.265 | ** | -0.088 | |
| | (0.048) | | (0.094) | | (0.051) | | (0.209) | | (0.101) | | (0.103) | | (0.062) | |
|   Tertiary | -0.095 | | -0.226 | | 0.038 | | -0.043 | | 0.269 | ** | -0.337 | * | -0.485 | ** |
| | (0.082) | | (0.140) | | (0.096) | | (0.250) | | (0.126) | | (0.190) | | (0.227) | |
| Marital status (reference: married) | | | | | | | | | | | | | | |
|   Never married | -0.040 | | -0.119 | | 0.030 | | -0.262 | | -0.048 | | -0.051 | | 0.052 | |
| | (0.086) | | (0.126) | | (0.114) | | (0.195) | | (0.172) | | (0.182) | | (0.147) | |
|   Widowed | 0.103 | *** | 0.200 | *** | 0.119 | *** | 0.177 | | 0.202 | *** | 0.149 | * | 0.085 | * |
| | (0.036) | | (0.073) | | (0.038) | | (0.148) | | (0.069) | | (0.088) | | (0.046) | |
|   Separated/ divorced | 0.000 | | -0.013 | | 0.078 | | 0.138 | | -0.127 | | -0.104 | | 0.134 | |
| | (0.074) | | (0.099) | | (0.109) | | (0.167) | | (0.181) | | (0.129) | | (0.128) | |
| Urban | -0.017 | | 0.010 | | -0.035 | | 0.007 | | -0.002 | | 0.006 | | -0.061 | ** |
| | (0.017) | | (0.029) | | (0.022) | | (0.051) | | (0.034) | | (0.035) | | (0.028) | |
| Peninsular Malaysia | -0.031 | | -0.217 | *** | 0.141 | *** | -0.045 | | 0.254 | *** | -0.338 | *** | 0.044 | |
| | (0.041) | | (0.064) | | (0.048) | | (0.111) | | (0.072) | | (0.080) | | (0.068) | |

**Table 2 (continued)**

| Variable/Statistic | All | | Aged 40–59 | | Aged 60+ | | Men Aged 40–59 | | Men Aged 60+ | | Women Aged 40–59 | | Women Aged 60+ | |
|---|---|---|---|---|---|---|---|---|---|---|---|---|---|---|
| Living children (reference: none) | | | | | | | | | | | | | | |
| 1–2 | 0.200 | *** | 0.045 | | 0.247 | ** | -0.110 | | 0.377 | *** | 0.162 | | 0.155 | |
| | (0.072) | | (0.105) | | (0.099) | | (0.171) | | (0.137) | | (0.140) | | (0.135) | |
| 3–4 | 0.209 | *** | 0.061 | | 0.260 | *** | -0.109 | | 0.323 | ** | 0.182 | | 0.226 | * |
| | (0.072) | | (0.109) | | (0.099) | | (0.181) | | (0.138) | | (0.143) | | (0.134) | |
| 5+ | 0.203 | *** | -0.001 | | 0.238 | ** | -0.051 | | 0.265 | * | 0.054 | | 0.228 | * |
| | (0.077) | | (0.121) | | (0.102) | | (0.201) | | (0.145) | | (0.156) | | (0.137) | |
| Living arrangements (reference: live alone) | | | | | | | | | | | | | | |
| Spouse only | 0.014 | | -0.004 | | 0.021 | | -0.138 | | 0.040 | | 0.091 | | -0.025 | |
| | (0.055) | | (0.119) | | (0.060) | | (0.171) | | (0.083) | | (0.174) | | (0.083) | |
| Other family members | -0.017 | | -0.115 | | 0.063 | | -0.136 | | 0.046 | | -0.082 | | 0.053 | |
| | (0.049) | | (0.107) | | (0.053) | | (0.138) | | (0.074) | | (0.165) | | (0.070) | |
| Nearest living son (reference: living elsewhere) | | | | | | | | | | | | | | |
| Living with | -0.054 | * | 0.071 | | -0.118 | *** | 0.086 | | -0.160 | *** | 0.084 | | -0.077 | |
| | (0.031) | | (0.053) | | (0.037) | | (0.089) | | (0.056) | | (0.067) | | (0.048) | |
| Leaving near | 0.032 | | 0.006 | | 0.051 | | 0.293 | * | 0.068 | | -0.169 | | 0.026 | |
| | (0.042) | | (0.116) | | (0.044) | | (0.169) | | (0.068) | | (0.157) | | (0.057) | |
| Nearest living daughter (reference: living elsewhere) | | | | | | | | | | | | | | |
| Living with | -0.022 | | -0.009 | | 0.032 | | 0.028 | | 0.065 | | -0.023 | | 0.017 | |
| | (0.031) | | (0.050) | | (0.037) | | (0.081) | | (0.057) | | (0.065) | | (0.049) | |
| Living near | 0.084 | ** | 0.343 | *** | 0.072 | * | 0.583 | *** | 0.105 | | 0.180 | * | 0.051 | |
| | (0.037) | | (0.083) | | (0.041) | | (0.145) | | (0.065) | | (0.103) | | (0.051) | |
| Second survey wave | 0.247 | *** | 0.242 | *** | 0.244 | *** | 0.287 | *** | 0.240 | *** | 0.219 | *** | 0.245 | *** |
| | (0.018) | | (0.029) | | (0.022) | | (0.052) | | (0.036) | | (0.035) | | (0.029) | |
| Constant | -2.167 | *** | -3.121 | *** | -0.343 | | -2.513 | *** | -0.854 | *** | -3.735 | *** | 0.135 | |
| | (0.148) | | (0.317) | | (0.225) | | (0.530) | | (0.329) | | (0.414) | | (0.297) | |
| No. of observations | 10,339 | | 5,976 | | 4,363 | | 2,561 | | 2,019 | | 3,415 | | 2,344 | |
| Model tests (p-values) | | | | | | | | | | | | | | |
| Model fit | 0.000 | | 0.000 | | 0.000 | | 0.000 | | 0.000 | | 0.000 | | 0.000 | |
| Random effects | 0.000 | | 0.000 | | 0.000 | | 0.000 | | 0.000 | | 0.000 | | 0.000 | |

*** = $p<0.01$, ** = $p<0.05$, and * = $p<0.10$.
Source: Authors' calculations using Malaysia Ageing and Retirement Survey data.

**Table 3. Poisson Random Effects Estimates for Number of Depressive Symptoms, Viet Nam**

| Variable/Statistic | All | | Aged 60+ | | All Men | | Men Aged 60+ | | All Women | | Women Aged 60+ | |
|---|---|---|---|---|---|---|---|---|---|---|---|---|
| Woman | 0.199 | *** | 0.154 | *** | | | | | | | | |
| | (0.000) | | (0.000) | | | | | | | | | |
| Age (years) | 0.007 | *** | 0.006 | *** | 0.014 | *** | 0.009 | *** | 0.003 | *** | 0.004 | *** |
| | (0.000) | | (0.000) | | (0.000) | | (0.000) | | (0.000) | | (0.000) | |
| Kinh ethnicity | -0.031 | *** | -0.014 | *** | 0.016 | *** | -0.032 | *** | -0.041 | *** | -0.015 | *** |
| | (0.000) | | (0.000) | | (0.000) | | (0.000) | | (0.000) | | (0.000) | |
| Education (reference: less than primary) | | | | | | | | | | | | |
| Primary | -0.074 | *** | -0.042 | *** | -0.032 | *** | -0.018 | *** | -0.110 | *** | -0.075 | *** |
| | (0.000) | | (0.000) | | (0.000) | | (0.000) | | (0.000) | | (0.000) | |
| Secondary | -0.262 | *** | -0.194 | *** | -0.219 | *** | -0.205 | *** | -0.294 | *** | -0.200 | *** |
| | (0.000) | | (0.000) | | (0.000) | | (0.000) | | (0.000) | | (0.000) | |
| Tertiary | -0.531 | *** | -0.488 | *** | -0.531 | *** | -0.467 | *** | -0.518 | *** | -0.486 | *** |
| | (0.000) | | (0.000) | | (0.000) | | (0.000) | | (0.000) | | (0.000) | |
| Marital status (reference: married) | | | | | | | | | | | | |
| Never married | -0.249 | *** | -0.215 | *** | 0.242 | *** | 0.335 | *** | -0.296 | *** | -0.279 | *** |
| | (0.000) | | (0.000) | | (0.000) | | (0.000) | | (0.000) | | (0.000) | |
| Widowed | 0.054 | *** | 0.055 | *** | -0.076 | *** | -0.047 | *** | 0.125 | *** | 0.103 | *** |
| | (0.000) | | (0.000) | | (0.000) | | (0.000) | | (0.000) | | (0.000) | |
| Separated/divorced | -0.022 | *** | 0.198 | *** | 0.171 | *** | 0.026 | *** | -0.064 | *** | 0.177 | *** |
| | (0.000) | | (0.000) | | (0.000) | | (0.000) | | (0.000) | | (0.000) | |
| Urban | -0.135 | *** | -0.137 | *** | -0.053 | *** | -0.068 | *** | -0.209 | *** | -0.207 | *** |
| | (0.000) | | (0.000) | | (0.000) | | (0.000) | | (0.000) | | (0.000) | |
| Region (reference: northern) | | | | | | | | | | | | |
| Central | 0.047 | *** | 0.029 | *** | 0.048 | *** | 0.014 | *** | 0.037 | *** | 0.036 | *** |
| | (0.000) | | (0.000) | | (0.000) | | (0.000) | | (0.000) | | (0.000) | |
| Southern | -0.031 | *** | -0.005 | *** | 0.029 | *** | 0.064 | *** | -0.080 | *** | -0.054 | *** |
| | (0.000) | | (0.000) | | (0.000) | | (0.000) | | (0.000) | | (0.000) | |
| Living children (reference: none) | | | | | | | | | | | | |
| 1–2 | -0.261 | *** | -0.174 | *** | -0.486 | *** | -0.450 | *** | -0.124 | *** | -0.048 | *** |
| | (0.000) | | (0.000) | | (0.000) | | (0.000) | | (0.000) | | (0.000) | |
| 3–4 | -0.199 | *** | -0.154 | *** | -0.236 | *** | -0.127 | *** | -0.163 | *** | -0.159 | *** |
| | (0.000) | | (0.000) | | (0.000) | | (0.000) | | (0.000) | | (0.000) | |
| 5+ | -0.142 | *** | -0.120 | *** | -0.165 | *** | -0.052 | *** | -0.138 | *** | -0.174 | *** |
| | (0.000) | | (0.000) | | (0.000) | | (0.000) | | (0.000) | | (0.000) | |

**Table 3 (continued)**

| Variable/Statistic | All | | Aged 60+ | | All Men | | Men Aged 60+ | | All Women | | Women Aged 60+ | |
|---|---|---|---|---|---|---|---|---|---|---|---|---|
| Living arrangements (reference: live alone) | | | | | | | | | | | | |
| Spouse only | -0.230 | *** | -0.179 | *** | -0.402 | *** | -0.349 | *** | -0.091 | *** | -0.072 | *** |
| | (0.000) | | (0.000) | | (0.000) | | (0.000) | | (0.000) | | (0.000) | |
| Other family members | -0.136 | *** | -0.078 | *** | -0.369 | *** | -0.301 | *** | -0.021 | *** | 0.017 | *** |
| | (0.000) | | (0.000) | | (0.000) | | (0.000) | | (0.000) | | (0.000) | |
| Nearest living child (reference: living elsewhere) | | | | | | | | | | | | |
| Living with | -0.101 | *** | -0.091 | *** | 0.075 | *** | 0.019 | *** | -0.205 | *** | -0.180 | *** |
| | (0.000) | | (0.000) | | (0.000) | | (0.000) | | (0.000) | | (0.000) | |
| Leaving near | -0.042 | *** | -0.039 | *** | 0.024 | *** | 0.001 | *** | -0.104 | *** | -0.092 | *** |
| | (0.000) | | (0.000) | | (0.000) | | (0.000) | | (0.000) | | (0.000) | |
| 2022 survey wave | -0.098 | *** | -0.121 | *** | -0.180 | *** | -0.210 | *** | -0.037 | *** | -0.052 | *** |
| | (0.000) | | (0.000) | | (0.000) | | (0.000) | | (0.000) | | (0.000) | |
| Constant | 1.317 | *** | 1.288 | *** | 0.897 | *** | 1.206 | *** | 1.777 | *** | 1.591 | *** |
| | (0.000) | | (0.000) | | (0.000) | | (0.000) | | (0.000) | | (0.000) | |
| | | | | | | | | | | | | |
| No. of observations | 7,192 | | 5,941 | | 2,935 | | 2,423 | | 4,257 | | 3,518 | |
| Model tests (p-values) | | | | | | | | | | | | |
| Model fit | 0.000 | | 0.000 | | 0.000 | | 0.000 | | 0.000 | | 0.000 | |
| Random effects | 0.000 | | 0.000 | | 0.000 | | 0.000 | | 0.000 | | 0.000 | |

*** = $p<0.01$, ** = $p<0.05$, and * = $p<0.10$.
Source: Authors' calculations using Viet Nam Aging Survey data.

**Table 4. Poisson Random Effects Estimates for Number of Chronic Health Conditions, Viet Nam**

| Variable/Statistic | All | | Aged 60+ | | All Men | | Men Aged 60+ | | All Women | | Women Aged 60+ | |
|---|---|---|---|---|---|---|---|---|---|---|---|---|
| Woman | 0.104 | *** | 0.109 | *** | | | | | | | | |
| | (0.000) | | (0.000) | | | | | | | | | |
| Age (years) | 0.009 | *** | 0.005 | *** | 0.014 | *** | 0.013 | *** | 0.005 | *** | 0.000 | *** |
| | (0.000) | | (0.000) | | (0.000) | | (0.000) | | (0.000) | | (0.000) | |
| Kinh ethnicity | 0.129 | *** | 0.068 | *** | 0.050 | *** | 0.052 | *** | 0.158 | *** | 0.077 | *** |
| | (0.000) | | (0.000) | | (0.000) | | (0.000) | | (0.000) | | (0.000) | |
| Education (reference: less than primary) | | | | | | | | | | | | |
| Primary | -0.039 | *** | -0.030 | *** | -0.142 | *** | -0.176 | *** | 0.032 | *** | 0.074 | *** |
| | (0.000) | | (0.000) | | (0.000) | | (0.000) | | (0.000) | | (0.000) | |
| Secondary | 0.132 | *** | 0.147 | *** | 0.159 | *** | 0.164 | *** | 0.097 | *** | 0.104 | *** |
| | (0.000) | | (0.000) | | (0.000) | | (0.000) | | (0.000) | | (0.000) | |
| Tertiary | 0.141 | *** | 0.127 | *** | 0.230 | *** | 0.201 | *** | 0.013 | *** | -0.016 | *** |
| | (0.000) | | (0.000) | | (0.000) | | (0.000) | | (0.000) | | (0.000) | |
| Marital status (reference: married) | | | | | | | | | | | | |
| Never married | -0.109 | *** | -0.174 | *** | 0.088 | *** | -0.096 | *** | -0.174 | *** | -0.171 | *** |
| | (0.000) | | (0.000) | | (0.000) | | (0.000) | | (0.000) | | (0.000) | |
| Widowed | -0.009 | *** | 0.018 | *** | -0.126 | *** | -0.100 | *** | 0.039 | *** | 0.070 | *** |
| | (0.000) | | (0.000) | | (0.000) | | (0.000) | | (0.000) | | (0.000) | |
| Separated/divorced | -0.087 | *** | 0.030 | *** | 0.066 | *** | -0.143 | *** | -0.114 | *** | 0.043 | *** |
| | (0.000) | | (0.000) | | (0.000) | | (0.000) | | (0.000) | | (0.000) | |
| Urban | 0.032 | *** | 0.062 | *** | 0.117 | *** | 0.144 | *** | -0.036 | *** | -0.002 | *** |
| | (0.000) | | (0.000) | | (0.000) | | (0.000) | | (0.000) | | (0.000) | |
| Region (reference: northern) | | | | | | | | | | | | |
| Central | -0.057 | *** | -0.053 | *** | -0.052 | *** | -0.055 | *** | -0.062 | *** | -0.045 | *** |
| | (0.000) | | (0.000) | | (0.000) | | (0.000) | | (0.000) | | (0.000) | |
| Southern | 0.140 | *** | 0.124 | *** | 0.078 | *** | 0.051 | *** | 0.186 | *** | 0.176 | *** |
| | (0.000) | | (0.000) | | (0.000) | | (0.000) | | (0.000) | | (0.000) | |
| Living children (reference: none) | | | | | | | | | | | | |
| 1–2 | 0.090 | *** | 0.040 | *** | 0.535 | *** | 0.090 | *** | -0.042 | *** | 0.050 | *** |
| | (0.000) | | (0.000) | | (0.000) | | (0.000) | | (0.000) | | (0.000) | |
| 3–4 | 0.125 | *** | 0.010 | *** | 0.605 | *** | 0.137 | *** | -0.047 | *** | -0.037 | *** |
| | (0.000) | | (0.000) | | (0.000) | | (0.000) | | (0.000) | | (0.000) | |
| 5+ | 0.305 | *** | 0.229 | *** | 0.781 | *** | 0.360 | *** | 0.122 | *** | 0.161 | *** |
| | (0.000) | | (0.000) | | (0.000) | | (0.000) | | (0.000) | | (0.000) | |

**Table 4 (continued)**

| Variable/Statistic | All | | Aged 60+ | | All Men | | Men Aged 60+ | | All Women | | Women Aged 60+ | |
|---|---|---|---|---|---|---|---|---|---|---|---|---|
| Living arrangements (reference: live alone) | | | | | | | | | | | | |
| Spouse only | -0.134 | *** | -0.047 | *** | -0.267 | *** | -0.175 | *** | -0.050 | *** | 0.029 | *** |
| | (0.000) | | (0.000) | | (0.000) | | (0.000) | | (0.000) | | (0.000) | |
| Other family members | -0.094 | *** | -0.023 | *** | -0.146 | *** | -0.051 | *** | -0.066 | *** | -0.018 | *** |
| | (0.000) | | (0.000) | | (0.000) | | (0.000) | | (0.000) | | (0.000) | |
| Nearest living child (reference: living elsewhere) | | | | | | | | | | | | |
| Living with | -0.097 | *** | -0.117 | *** | -0.169 | *** | -0.194 | *** | -0.042 | *** | -0.069 | *** |
| | (0.000) | | (0.000) | | (0.000) | | (0.000) | | (0.000) | | (0.000) | |
| Leaving near | -0.035 | *** | -0.064 | *** | -0.072 | *** | -0.097 | *** | 0.011 | *** | -0.032 | *** |
| | (0.000) | | (0.000) | | (0.000) | | (0.000) | | (0.000) | | (0.000) | |
| 2022 survey wave | 0.489 | *** | 0.466 | *** | 0.434 | *** | 0.441 | *** | 0.521 | *** | 0.477 | *** |
| | (0.000) | | (0.000) | | (0.000) | | (0.000) | | (0.000) | | (0.000) | |
| Constant | -0.456 | *** | -0.117 | *** | -1.054 | *** | -0.606 | *** | -0.051 | *** | 0.346 | *** |
| | (0.000) | | (0.000) | | (0.000) | | (0.000) | | (0.000) | | (0.000) | |
| No. of observations | 7,498 | | 6,239 | | 3,035 | | 2,519 | | 4,463 | | 3,720 | |
| Model tests (p-values) | | | | | | | | | | | | |
| Model fit | 0.000 | | 0.000 | | 0.000 | | 0.000 | | 0.000 | | 0.000 | |
| Random effects | 0.000 | | 0.000 | | 0.000 | | 0.000 | | 0.000 | | 0.000 | |

Notes: The notation *** is p<0.01, ** is p<0.05, and * is p<0.10.
Source: Authors' calculations using Viet Nam Aging Survey data.